\documentclass[twocolumn, pra,showpacs,nofootinbib]{revtex4}

\usepackage{dcolumn,graphicx}
\begin{document}
\preprint{UNR Feb 2002-\today }
\title{Ultracold collision properties of metastable alkaline-earth atoms}
\author{Andrei Derevianko and Sergey G. Porsev}
\affiliation{Department of Physics, University of Nevada, Reno, Nevada 89557}
\author{Svetlana Kotochigova, Eite Tiesinga, and Paul S. Julienne}
\affiliation {Atomic physics division, National Institute of Standards and
 Technology, 100 Bureau Drive, stop 8423, Gaithersburg, Maryland 20899}
\date{\today}
\begin{abstract}
Ultra-cold collisions of spin-polarized
$^{24}\mathrm{Mg}$,$^{40}\mathrm{Ca}$, and $^{88}\mathrm{Sr}$ in the
metastable $^3\!P_2$ excited state are investigated.  We calculate
the long-range interaction potentials and estimate the scattering
length and the collisional loss rate as a function of magnetic field.
The estimates are based on molecular potentials between $^3\!P_2$
alkaline-earth atoms obtained from {\it ab initio} atomic and
molecular structure calculations.  The scattering lengths show resonance
behavior due to the appearance of a molecular bound state in a purely
long-range interaction potential and are positive for magnetic fields
below 50 mT.  A loss-rate model shows that losses should be
smallest near zero magnetic field and for fields slightly larger than
the resonance field, where the scattering length is also positive.
\end{abstract}

\maketitle

Exploring ultracold collision physics with metastable $(nsnp)^3\!P_2$
alkaline-earth atoms \cite{Der01,KatIdoIso00,Kil02,LofXuHal02,GruHem02}
promises insights that complement those obtained from the more
conventional atomic species used in laser cooling. Unlike alkali-metal
atoms \cite{Burnett02,AlkaliMetal} the most-common alkaline-earth
atoms have no nuclear spin, which greatly simplifies a theoretical
description\cite{MacJulSuo01}.  Furthermore, in contrast to metastable
noble-gas collisions\cite{nobleGasesAll} the low electronic energy of
the alkaline-earth metastable states ensures the absence of collisionally
induced Penning or associative ionization.

Atomic collisions in the ultracold regime play a crucial role on the road
toward quantum degenerate gases.  We show here that polarized metastable
$(nsnp)^3\!P_2$ alkaline earth systems with projection $m=2$ along the
magnetic field might satisfy the key requirements for this quest: a
positive scattering length and a favorable ratio of elastic to inelastic
collision rates for certain magnetic field strengths.  In addition, we
show how a new type of pure long range molecular states can allow the
resonant magnetic field control of the scattering length.  These states
arise due to an interplay of an anisotropic (quadrupole) interaction and
the magnetic field.  Similar anisotropy has been predicted to exist for
polar molecule collisions in an electric field\cite{AvdBoh02}.  Many-body
systems with anistropic interactions might now be explored\cite{You02}.

Trapping of metastable alkaline-earths in magneto-optical
\cite{KatIdoIso00,Kil02,LofXuHal02,GruHem02} and magnetic traps
\cite{KatIdoIso00,LofBocMos02,Kil02,LofXuHal02} has been demonstrated.
A $\sim$1 s magnetic trap lifetime for densities near $10^{10}$
atoms/cm$^3$ is observed with metastable $^{88}$Sr in experiments at
Rice and Tokyo university\cite{Kil02,Tokyo}.  These trap lifetimes are
limited by background collisions rather than by the radiative lifetime,
which is on the order of minutes\cite{Der01}.

{\it Molecular potentials ---} We have calculated short-range
adiabatic potentials correlating to two $^3$P$_2$ atoms using a
molecular {\it ab initio} relativistic valence bond method, as previously
described\cite{Kotot02}. A variety of attractive and repulsive potentials,
and a number of short-range avoided crossings appear among them.
Strong collisional loss processes are likely~\cite{MacJulSuo01} if the
atoms approach one another on the attractive curves.  The section
below shows how a magnetic field can be used to manipulate long-range
properties in such a way as to create pure long-range molecular states
and prevent some of these losses.

Atomic structure calculations and Rayleigh-Schr\"{o}dinger perturbation
theory allow us to express the long-range adiabatic potentials in terms
of multipole moments of the atoms\cite{DalDav66}.  The matrix elements
of the Hamiltonian for two $^3\!P_2$ atoms up to order $1/R^6$, where $R$
is the internuclear separation, are \cite{DerDal00}
\begin{eqnarray}
  \langle x | H_{lr} | y \rangle &=&
   \frac{\langle x|V_{QQ}|y \rangle}{R^5}
    + \sum_{j,j'=1}^3 \langle x|A_{jj'}|y\rangle  \frac{C_6^{jj'}}{R^6} \,.
\label{Heff2_Eqn}
\end{eqnarray}
Here $|x\rangle$ and $|y\rangle$ are product states $|^3\!P_2, \Omega_1
\rangle|^3\!P_2, \Omega_2 \rangle$, $\Omega_\alpha$ is the projection of
the total electron angular momentum $j_\alpha=2$ of atom $\alpha$ on the
internuclear axis, and $\Omega=\Omega_1+\Omega_2$, which is a good quantum
number in the Hund's case (c) coupling scheme.  The two terms on the right
hand side of Eq.~\ref{Heff2_Eqn} describe the quadrupole-quadrupole (QQ)
and the second-order dipole-dipole (DD) interaction.

Matrix elements of the QQ interaction are expressed in terms of the
quadrupole moment $\mathcal{Q}$ of the $^3\!P_2$ state\cite{Der01}.
For this paper the value of $\mathcal{Q}$ is recalculated using a more
accurate many-body technique\cite{PorKozRak01}.  The values are 8.46(8)
$E_h a_0^5$, 12.9(4) $E_h a_0^5$, and 15.6(5) $E_h a_0^5$ for metastable
Mg, Ca, and Sr, respectively.  Here $E_h$ is a Hartree, 1 $a_0$ is 0.0529
nm, and one-standard deviation uncertainties, based on convergence studies
of the many-body theory, are given in parenthesis.  The $C^{jj'}_6$
are intermediate dipole-dipole dispersion coefficients, calculated
following Ref.~\cite{PorDer02}, and tabulated in Table~\ref{Tbl_intC6}.
The $\langle x|A_{jj'}|y\rangle$ only depend on angular momentum algebra
and are given in Ref.~\cite{DerDal00}.

\begin{table}
\caption{Intermediate dispersion coefficients
${C}_6^{jj'}$ in units of $10^3\,E_h a_0^6$. The $C_6$ are
symmetric in $j$ and $j'$and have a 10\% one-standard deviation
uncertainty.}
\label{Tbl_intC6}
\begin{ruledtabular}
\begin{tabular}{ccccccc}
   & $C_6^{11}$ & $C_6^{21}$ & $C_6^{22}$ & $C_6^{31}$ &  $C_6^{32}$
   & $C_6^{33}$\\
\hline
 Mg$^*$  &  3.19     &  $-$3.70  &   4.40  &   6.47  &  $-$7.60  &  13.2  \\
 Ca$^*$  &  7.74     & $-$10.4  &  14.1  &  19.0  & $-$25.8  &  50.6  \\
 Sr$^*$  & 13.3     & $-$17.1  &  22.3  &  35.0  & $-$46.8  & 109
\end{tabular}
\end{ruledtabular}
\end{table}

The Hund's case(c) adiabatic potentials that connect to our short-range
valence-bond calculation are obtained by diagonalizing Eq.~\ref{Heff2_Eqn}
within the $|^3\!P_2, \Omega_1 \rangle|^3\!P_2, \Omega_2 \rangle$
basis (See for examle Ref.~\cite{Der01}.).

{\it Magnetic field ---} A magnetic field $\mathbf{B}$ modifies
the long-range interaction of two metastable alkaline-earth atoms profoundly.
For each atom $\alpha$ the approximate Zeeman Hamiltonian $ H_{Z\alpha}=
\mu_{B}\left(  \mathbf{j}_\alpha
                       +\mathbf{s}_\alpha\right) \cdot \mathbf{B} $
has been added to the other terms in the molecular Hamiltonian.  Here,
$\mathbf{s}_\alpha$ is the electron spin of atom $\alpha$, $\mu_{B}$
is the Bohr magneton, and we have assumed that the atomic gyromagnetic
ratio of the electron orbital angular momentum and spin are one and two,
respectively.  The magnetic field lifts the degeneracy with respect to
the projection $m_\alpha$ of $\mathbf{j}_\alpha$ along the space-fixed
magnetic field direction.  Forces due to the interatomic interaction break
this space-fixed quantization, and align the molecular angular momentum
along the internuclear axis.  This results in loss of polarization of
the angular momentum.

\begin{figure}[ht] \begin{center}
\includegraphics*[scale=0.75]{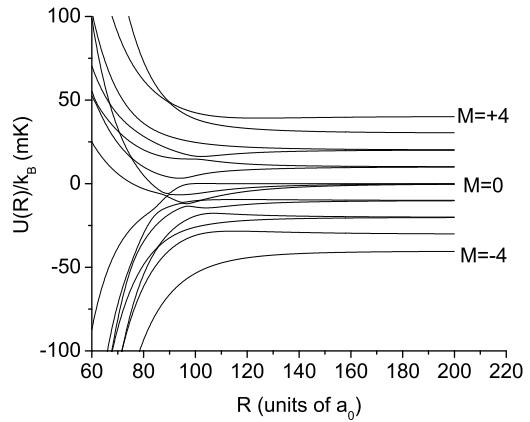} \caption{Long-range
gerade adiabatic potentials correlating to two $^3\!P_2$ Sr atoms for $B=10$
mT and $\theta_B=45^\circ$.
\label{Fig_Rdep_45}} \end{center}
\end{figure}

The long-range adiabatic molecular potentials in the presence of a
magnetic field are found by diagonalizing $H_{Z1}+H_{Z2}+H_{lr}$
within the $^3\!P_2$+$^3\!P_2$ product basis. These potentials
$U(R,\theta_B)$ not only depend on the field strength but also on
the angle $\theta_B$ between the internuclear axis and $\mathbf{B}$.
Figure~\ref{Fig_Rdep_45} shows all Sr$_2$ gerade adiabatic potentials
for a 10 mT field and $\theta_B=45^\circ$.  The nine dissociation limits
are separated by about $E/k_B=$10 mK and are labeled by $M=m_1+m_2$,
where $M=+4$($-4$) for the highest(lowest) dissociation limit. $k_B$ is the
Boltzmann constant.

Figure~\ref{Fig_Rdep_45} shows a multitude of avoided crossings for
internuclear separations where the Zeeman splitting is comparable to
the quadrupole-quadrupole interactions, that is, avoided crossings appear for
$R \approx R_B\equiv\sqrt[5]{\mathcal{Q}^2/(\mu_BB)}$.  For Sr$^{\ast}$
and $B=10$ mT,  $R_B$ is $\sim100$ $a_0$.  For $R<R_B$ the QQ
interactions dominate and the Hund's case (c) $\Omega_g^\pm$ molecular
potentials correlating to two $^3\!P_2$ atoms are recovered.

\begin{figure}[h] \begin{center}
\includegraphics*[scale=0.7]{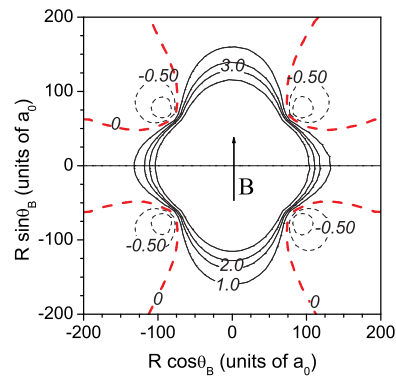} \caption{Polar plot of the
adiabatic potential for two spin polarized  Sr$^*$ atoms at $B$=10 mT.
Equipotential curves are shown as dashed, thick dashed, and solid
contours for energies that are lower than, equal to, and higher than
the dissociation energy of the two atoms, respectively.  The contour values
are given in mK.  The arrow indicates the direction of ${\mathbf B}$.
\label{Fig_contour}}
\end{center} \end{figure}

Experiments trap the ``low-field seeking'' spin-polarized $m_\alpha=+2$
state \cite{Kil02,LofXuHal02}.  The top-most potential in
Fig.~\ref{Fig_Rdep_45} dissociates to this limit. The radial and angular
dependence of this potential $U_{m_1=2,m_2=2}(R,\theta_B)$ is shown in
Fig.~\ref{Fig_contour}.  Four long-range hills and valleys are visible
for $R>R_B$ and a repulsive hard core that is nearly independent of
angle appears for $R\approx R_B$.

\begin{figure}[h] \begin{center}
\includegraphics*[scale=0.75]{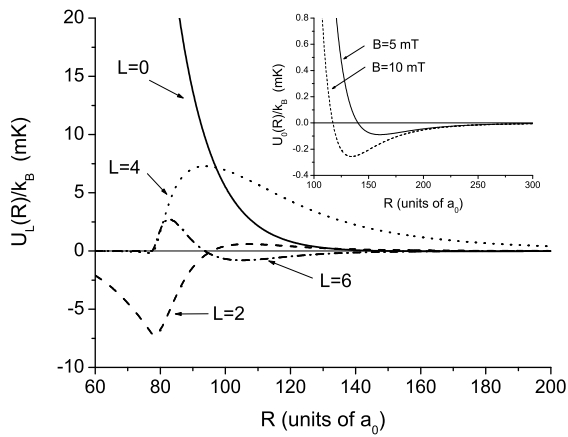}
\caption{The coefficients $U_L(R)$ of Eq.~\protect\ref{Eqn_expand}
for Sr$_2$ as a function of internuclear separation and $B=$10 mT.
The inset shows $U_0(R)$ for $B=5$ mT and 10 mT.
\label{Fig_expansion}}
\end{center}
\end{figure}

The behavior of $U_{2,2}(R,\theta_B)$ is further illuminated when it
is expanded in terms of Legendre polynomials $P_L(x)$,
\begin{equation}
U_{2,2}\left( R, \theta_B \right) = \sum_{L=0,2,4,\ldots} U_L(R)\,P_L(x)
\, , \label{Eqn_expand}
\end{equation}
where $x=\cos\theta_B$.  The $U_L(R)$ of Sr$_2$ are shown in
Fig.~\ref{Fig_expansion}.  For $R \gg R_B$, $U_4$ is largest and is
dominated by the QQ interaction.  In Fig.~\ref{Fig_contour} it gives
rise to the hills and valleys.  For $R\approx R_B$, contributions of
other components are appreciable, while for $R \lesssim \, 60 \, a_0$
the repulsive $U_0(R)$ dominates. The inset of Fig.~\ref{Fig_expansion}
shows that $U_0(R)$ has a shallow attractive region for large $R$, that
becomes deeper and wider for increasing $B$. This so-called ``Zeeman-van
der Waals'' well is a consequence of an interplay between Zeeman and
anisotropic quadrupole-quadrupole interactions.

For $R \gg R_B$, an analytical expression for $U_{2,2}$
can be derived based on perturbation theory around $|^3\!P_2,
m_1 \rangle|^3\!P_2, m_2\rangle$ with the quantization axis along
$\mathbf{B}$. We find for the contribution of the QQ interaction
\begin{equation}
U_{2,2}^{(QQ)}=\frac{3}{2}\frac{\mathcal{Q}^{2}}{R^{5} }\,P_4( x )
\,, \label{Eqn_ZeeLR}
\end{equation}
and for the DD interaction
\begin{eqnarray}
U_{2,2}^{(DD)} &=& - \frac{1}{R^6}
                   \left[ C_6^{(0)} +
                            C_6^{(2)}\,P_2( x ) +
                              C_6^{(4)}\,P_4( x )\right]\,.
\label{Eqn_C6expand}
\end{eqnarray}
The monopole coefficient $C_6^{(0)}$ is independent of $B$ and
is 1000  $E_ha_0^6$, 3300 $E_ha_0^6$, and 6200 $E_ha_0^6$ for
$\mathrm{Mg}^{\ast}$, $\mathrm{Ca}^{\ast}$, and $\mathrm{Sr}^{\ast}$
respectively.  Consequently, the long-range behavior of $U_0(R)$ is
given by an attractive $1/R^6$ potential that is independent of $B$.

{\it Scattering lengths ---} A rigorous description of elastic collisions
between spin polarized metastable $^3\!P\, j=2,m=2$ atoms should start
from $B=0$ theories\cite{collTheoryAll} and extend those to
include the Zeeman interaction\cite{AlkaliMetal}. Such a model is beyond
the scope of this paper. Instead, we take advantage of our understanding
of the adiabatic potentials in a magnetic field. Nuclear motion couples
the electronic potentials via so-called non-adiabatic couplings. These
couplings are strongest near avoided crossings between adiabatic
electronic potentials and lead to inelastic losses.  For example,
a transition from the $M=4$ to $M=3$ curves in Fig.~\ref{Fig_Rdep_45}
results in depolarization and conversion of internal energy to kinetic
energy.

\begin{figure}[h] \begin{center}
\includegraphics*[scale=0.75]{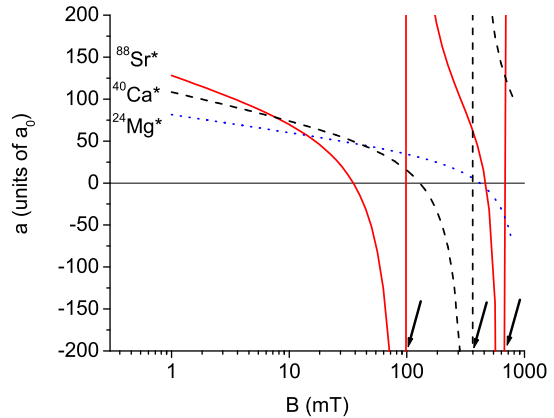}
\caption{Scattering length of the $U_0(R)$ potential
of spin polarized $^3\!P_2$ atoms as
a function of magnetic field. Values for $^{24}$Mg, $^{40}$Ca, and
$^{88}$Sr are shown.  The arrows show the magnetic fields for which
zero-energy resonances occur.
\label{Fig_scatLen}}
\end{center}
\end{figure}

First we will assume that adiabaticity holds. We relax this assumption
below. The zero-energy $s$-wave ($l=0$) scattering length of polarized
metastable $^3P_2$ $m=2$ atoms is then given by the scattering length
of the $U_0(R)$ component of $U_{2,2}\left( R, \theta_B \right)$.
This length for three alkaline-earth species as a function of magnetic
field is shown in Fig.~\ref{Fig_scatLen}.  For $B<40$ mT the scattering
lengths are large and positive and a Bose condensate would be stable
against collapse.  Uncertainties in QQ and DD coefficients do not change
the picture significantly.

For $B>$ 70 mT the scattering lengths exhibit singularities, where $a$
is infinite. These resonances are due the appearance of the first bound
state in $U_0(R)$.  The inset of Fig.~\ref{Fig_expansion} shows that
for increasing $B$ the well becomes wider and deeper, supporting an
increasingly larger number of bound states and thus extra resonances.

Magnetic field control of the sign and magnitude of the scattering
length near the resonances appears likely for these systems.  Magnetic
field-induced resonance behavior in scattering lengths has been observed
for alkali-metal atoms.  There, however, the singularity is a consequence
of Feshbach resonances rather than through the appearance of extra bound
states in an adiabatic potential.

{\em Inelastic processes ---} The scattering lengths have been determined
assuming that nonadiabatic couplings and thus losses are negligible.
We verify the applicability of the adiabatic approximation by estimating
loss rates from a two-channel curve-crossing model. The model has an
attractive van-der-Waals potential in an incoming channel, $|i\rangle$,
of two spin polarized atoms and $l=0$ that crosses a repulsive $C_5/R^5$
exit channel, $|e\rangle$, of two atoms with $m_1+m_2<4$ and $l=4$.
The coupling, $C'_5/R^5$, is due to the QQ interaction.  More precisely,
\begin{eqnarray}
  V_{\rm 2ch} &= &
                \left(
                    \begin{array}{cc}
                         -C_6^{(0)}/R^6    &     C'_5/R^5 \\
                         C'_5/R^5  &
                       C_5/R^5 +
              20\hbar^2/2\mu R^2 - \Delta
                    \end{array}
                \right) \,,
\end{eqnarray}
where $\mu$ is the reduced mass of the dimer, $\Delta \propto B$
is the difference in dissociation energy of the two states, the
exit state has a $l=4$ centrifugal potential, and  $C_6^{(0)}$
is defined in Eq.~\ref{Eqn_C6expand}.  The two states cross
at near $R_B$.

The $R$-dependent eigenvalues of $V_{\rm 2ch}$ are our model adiabatic
potentials. In fact, the top-most adiabat plays the role of the $L=0$
component of the $U_{22}$ adiabat. It has an attractive van der Waals
tail and a repulsive wall that is to be compared to the ``Zeeman-van
der Waals'' well of Fig.~\ref{Fig_expansion}.

Loss can occur when the atoms reach short-range $R$ on the incoming
channel or when the atoms are reflected into the repulsive exit channel.
We numerically solved the two coupled Schr\"odinger equations assuming
a hard-wall placed inside $R_B$ and parameter values valid for
strontium.  This approach has no losses due to short-range inelastic
processes. Nevertheless, we feel that the model gives an order of
magnitude estimate of the loss rate.  The calculations show that for $B$
from 1 mT to 5 mT the loss rate $K$ into the exit channel rapidly rises
from $\sim 10^{-13}$ cm$^3$/s to $10^{-11}$ cm$^3$/s due
to Wigner-threshold effects in the exit channel. For large fields the
rate decreases slowly and exponentially with $B$ and for $B\approx  500$
mT has dropped to $10^{-12}$ cm$^3$/s.

The loss rate and the scattering length show a resonance similar to the
resonances shown in Fig.~\ref{Fig_scatLen} when an extra bound state
appears in the upper adiabat.  For collision energies below 1 $\mu$K the
loss rate approaches zero in a narrow $\approx 1$ mT region above the
resonance location where the scattering length is large and positive.
For collision energies above 100 $\mu$K the resonance is broadened
and will be hard to observe.  The position of the resonance and the
corresponding loss rate will change when all coupled channels are
included. However, the resonance and a (partial) drop in the loss rate
is expected to survive.

Consequently, we have identified magnetic field regions where losses are
smaller than or on the order of $10^{-13}$ cm$^3$/s, i.e.  for $B<1$ mT
and for magnetic field strengths just above the resonance.  For several
Tesla fields the loss rate will be small as well.  At a rate constant
of $10^{-13}$ cm$^3$/s the imaginary part of the scattering length
$hK/(2\mu)$\cite{ForDalXX}, is orders of magnitude smaller than the
real part of the scattering length shown in Fig.~\ref{Fig_scatLen}.
The collision is adiabatic for these field regions and the scattering
length is meaningful. Moreover, these rates lead to sample lifetimes of
1 s at densities of $10^{13}$ cm$^{-3}$.

{\em Conclusion ---}
We have calculated the interaction potentials of metastable
alkaline-earth dimers in the presence of a magnetic field.  A purely
long-range interaction potential is found to determine the scattering
length between two spin polarized $^3\!P_2$ alkaline-earth atoms. A
magnetic-field-induced resonance is observed, where the interaction
potential supports an extra bound state.  Reference~\cite{AvdBoh02}
predicts similar long-range states for polar molecules in an electric
field.  Existence of such molecular states should be a general property
of colliding atoms or molecules with an anisotropic interaction potential
in an external field that split a degeneracy.

Although full-scale multi-channel calculations of magnetic trap losses
and scattering lengths are desirable, a two channel model indicates that
the inelastic loss rate of the collision between two spin polarized atoms
are small for small magnetic field strengths and above the field-induced
resonance, where the scattering length is positive.  We uncovered several
unique features of collisions between metastable alkaline-earths, which
may offer new insights into the physics of ultracold quantum gasses.

This work was supported in part by the National Science Foundation
and the office of Naval Research.

\bibliography{scat3P2_add}

\end{document}